\documentclass[preprint,number,12pt,sort&compress]{elsarticle}

\usepackage{amssymb}

\usepackage{array}
\newcommand{\PreserveBackslash}[1]{\let\temp=\\#1\let\\=\temp}
\newcolumntype{C}[1]{>{\PreserveBackslash\centering}p{#1}}
\newcolumntype{R}[1]{>{\PreserveBackslash\raggedleft}p{#1}}
\newcolumntype{L}[1]{>{\PreserveBackslash\raggedright}p{#1}}
\newcommand\Rey{\mbox{\textit{Re}}}  

\emergencystretch=\maxdimen
\begin{document}

\begin{frontmatter}


\title{On the intrinsic three-dimensionality of the flow normal to a circular disk}

\author[label1,label2]{Xinliang Tian\corref{cor1}}
\ead{tianxinliang@sjtu.edu.cn}

\cortext[cor1]{Corresponding author}
\address[label1]{State Key Laboratory of Ocean Engineering, Shanghai Jiao Tong University, Shanghai 200240, China}
\address[label2]{Collaborative Innovation Center for Advanced Ship and Deep-Sea Exploration, Shanghai 200240, China}




\begin{abstract}
  Direct numerical simulations are performed for the steady flow normal to a circular disk at the Reynolds number of 1000. Numerical simulations are conducted with different levels of simplification procedure by reducing the azimuthal extension of the disk. The full-disk, the half-disk, the quarter-disk, the eighth-disk and the two-dimensional (2D) cases with the identical grid resolution are considered. Intrinsic three-dimensionality is identified in the wake of the circular disk. Both of the instantaneous and mean flow quantities are influenced by the simplification level significantly. The mean drag coefficient obtained from the 2D case is about only $36\%$ of that obtained from the three-dimensional (3D) simulation for the full-disk.
\end{abstract}

\end{frontmatter}


Computational fluid dynamics (CFD) method is widely recognized and used in the scientific research field of fluid dynamics. Accurate resolution of the three-dimensional flow requires extensive computing resources, and the computing cost increases dramatically as the flow becomes complicated.  To save the computing resources and speed up the simulations, it is usually necessary to simplify the computation with proper techniques. One of the most common techniques is the two-dimensional (2D) simplification for simulating the flow around cylindrical structures, even though the flow is three-dimensional (3D) indeed. However, this may not always work well because the 3D flow is not resolved at all in the 2D simulations. \citet{najjar1995effects} compared the results of the 2D and 3D simulations for the flow normal to a flat plate with an infinite span. The time-averaged drag coefficient was significantly overestimated in the 2D simulation compared with the results of the 3D simulation and the experimental data \citep{fage1927flow}. Thus, the evaluation on the validity of the simplification procedure used in CFD is an important issue in the fluid dynamics.

Similar problem may exist for the flow around an axis-symmetric body, because the axis-symmetric flow configuration could also be simplified to a 2D configuration based on the axis-symmetry assumption. It is known that the intrinsic three-dimensionality of the flow around an axis-symmetric body occurs when the Reynolds number exceeds some critical values. The critical Reynolds number is dependent on the shape of the body and normally below 300, e.g., $\Rey\approx210$ for a sphere in \citep{johnson1999flow,fabre2008bifurcations} and $\Rey\approx135$ for a circular disk in \citep{shenoy2008flow}. Although the simplified 2D simulation is not able to capture the 3D flow feature, this simplification procedure has been used in some numerical simulations, see e.g., the CFD studies in \citep{michael1966steady,rimon1969numerical,natarajan1993instability,Tao2003a,Tao2003}. To our best knowledge, few study has documented on the influence of the 2D assumption on the axis-symmetric flow yet.

\begin{figure}
\begin{center}
\begin{tabular}{ccccc}
\includegraphics[angle=0, height=0.23\columnwidth]{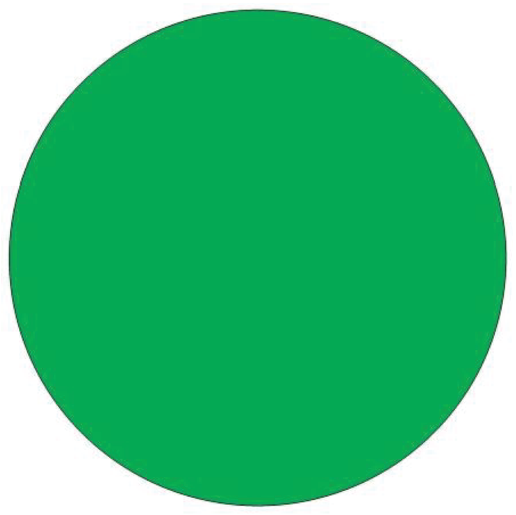} &
\includegraphics[angle=0, height=0.23\columnwidth]{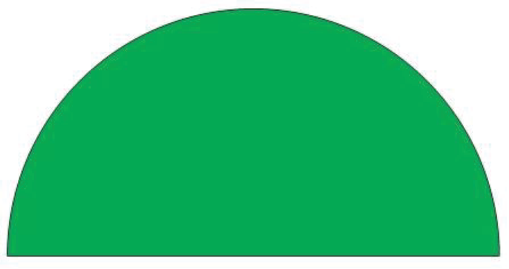} &
\includegraphics[angle=0, height=0.23\columnwidth]{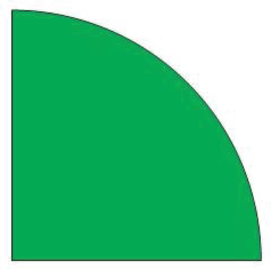} &
\includegraphics[angle=0, height=0.23\columnwidth]{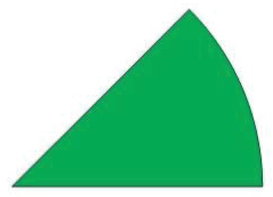} &
\includegraphics[angle=0, height=0.23\columnwidth]{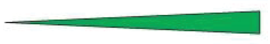}\\
$(a)$ & $(b)$ & $(c)$ & $(d)$ & $(e)$ \\
\end{tabular}
\end{center}
\caption{Overview of the disks with different simplification levels: $(a)$ the full-disk, $(b)$ the half-disk, $(c)$ the quarter-disk, $(d)$ the eighth-disk and $(e)$ the 2D case.}
\label{fig:disksoverview}
\end{figure}

In this letter, direct numerical simulations (DNS) for the flow around a circular disk are carried out. The Reynolds number is $\Rey=U_\infty D/\nu=1000$, where $U_\infty$ is the free stream velocity, $D$ is the diameter of the disk and $\nu$ is the kinematic viscosity of the fluid. For this nominally axis-symmetric flow configuration, simulations could be simplified by reducing the azimuthal extension of the disk. As shown in Fig.~\ref{fig:disksoverview}, five cases in total are carried out, in which four cases are the 3D simulations for the full-disk, the half-disk, the quarter-disk and the eighth-disk, respectively, and the rest one is the 2D simulation based on the axis-symmetric assumption. From the full-disk to the 2D case, the simplification level increases gradually and the computing cost decreases accordingly. The thickness ratio of the disk is $\chi=t_d/D=0.02$, where $t_d$ is the thickness of the disk.


The Cartesian coordinate system ($x,y,z$) is used in this study. If these coordinates are written as $(x_1, x_2, x_3)$ and the velocity component in the $x_i$-direction is denoted as $u_i$, where $i=1$--3, the Navier-Stokes (N-S) equations for an incompressible viscous fluid are written as follows:

\begin{equation}
\frac{\partial u_i}{\partial x_i}=0
\label{eq:NScontinuity}
\end{equation}

\begin{equation}
\frac{\partial u_i}{\partial t}+u_j \frac{\partial u_i}{\partial x_j} = -\frac{1}{\rho}\frac{\partial p}{\partial x_i}+\nu\frac{\partial^2 u_i}{\partial x_j \partial x_j}
\label{eq:NSMomentum}
\end{equation}
where $p$ is the pressure and $\rho$ is the density of the fluid.

The N-S equations are discretized using the finite volume method (FVM) based on the open source CFD code $OpenFOAM$. $OpenFOAM$ is mainly applied to solve problems in continuum mechanics. It is based on the tensorial approach and object oriented techniques \citep{weller1998tensorial}. The PISO (Pressure Implicit with Splitting of Operators) algorithm is used in this study. The spatial schemes for the interpolation, the gradient, the Laplacian and the divergence are linear, Gaussian linear, Gaussian linear corrected and Gaussian linear schemes, respectively. All of these schemes are in second order. The second order Crank-Nicolson scheme is used for the time integration. Further details of these schemes are given in OpenFOAM \citep{OpenFOAM2009}. The present numerical approach has already been applied successfully to simulate the incompressible flow around a circular disk \citep{Yang2014DiskInCurrent}.

\begin{figure}
\begin{center}
\begin{tabular}{c}
\includegraphics[angle=0, width=0.75\columnwidth]{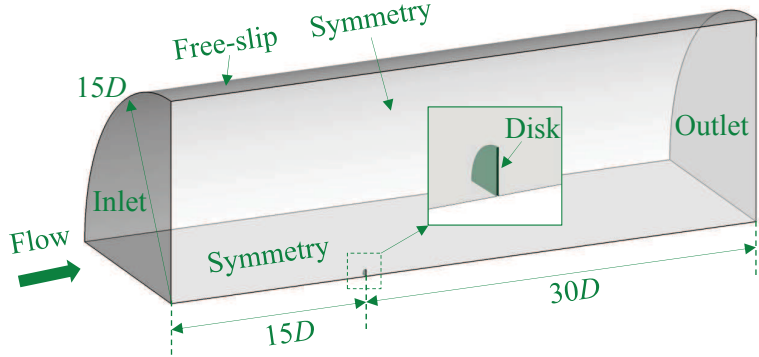}\\
\end{tabular}
\end{center}
\caption{The computational domain and the boundary conditions for the quarter-disk.}
\label{fig:DomainAndBoundary}
\end{figure}

The cylindrical computational domain is used in this study, see the example case for the quarter-disk shown in Fig.~\ref{fig:DomainAndBoundary}. The radius of the transverse cross-section of the computational domain is $15D$. The inlet and outlet boundaries are located at $15D$ upstream and $30D$ downstream to the center of the disk, respectively. This computational domain is much larger than the domain used in \citep{shenoy2008flow}, and therefore is considered to be large enough to eliminate the boundary effects.

On the disk surface, no-slip and zero normal pressure gradient boundary conditions are employed. At the inlet boundary, a uniform velocity $U_\infty$ and zero normal pressure gradient boundary conditions are prescribed. At the outlet boundary, the velocity is set to a zero normal gradient boundary condition, and the pressure is fixed to zero. On the side-wall of the computational domain, free-slip and zero normal gradient boundary conditions are applied for the velocity and the pressure, respectively. For the cases of the half-disk, the quarter-disk and the eighth-disk, the two planes in the azimuth direction are treated as the symmetry boundary condition. For the 2D case, there is only one mesh element in the azimuth direction, and the two planes in azimuth direction are treated as the "wedge" boundary condition. The "wedge" boundary condition is a specially designed boundary condition used for the axis-symmetry problem in $OpenFOAM$ .

The whole computational domain is discretized with hexahedral elements, and the grids near the surface of the disk and in the near wake are refined in order to resolve the steeper gradient there. The grid independency study has been carried out for the full-disk configuration, and three cases respectively with 2,129,088(coarse), 7,185,672(medium) and 17,032,704(fine) mesh elements are considered. The time-averaged value and the r.m.s.~(root-mean-square) of the drag coefficient are calculated over a period of $200D/U_\infty$ after the initial transient time. Comparison shows that the differences between the results of the last two cases are not significant. The results of the full-disk presented below are all based on the fine mesh case. In order to obtain an accurate prediction, the simulations for the the half-disk, the quarter-disk, the eighth-disk and the 2D case are all carried out with equal mesh resolution as the fine mesh used in the full-disk case, i.e., the mesh of 17,032,704 elements. The element size in the normal-wall direction next to the disk surface is $0.002D$. The maximum CFL (Courant-Friedrichs-Lewy) number in this study is always below 1. All the statistical quantities presented following are obtained from the simulations with a duration of $200D/U_\infty$ after the initial transient time.


\begin{figure}
\begin{center}
\begin{tabular}{c}
\includegraphics[angle=0, width=0.7\columnwidth]{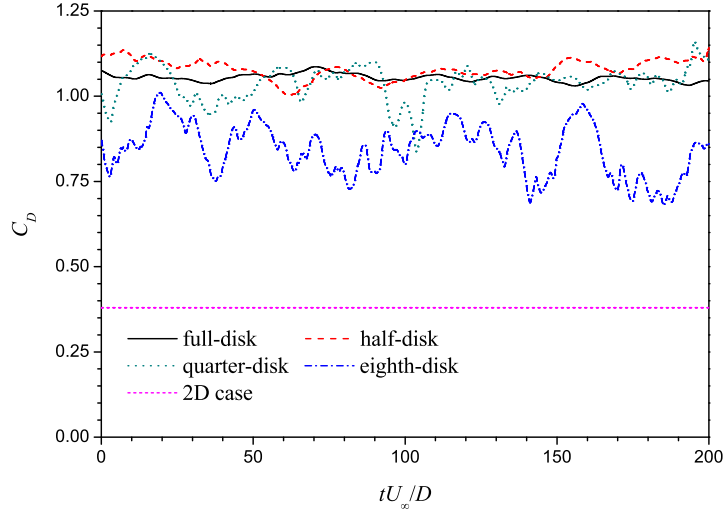}\\
\end{tabular}
\end{center}
\caption{Time-dependent variations of the drag coefficient.}
\label{fig:cdhistories}
\end{figure}

Fig.~\ref{fig:cdhistories} shows the time-dependent variations of the drag force coefficient $C_D$ obtained from the five cases with different simplification levels. The results of the initial transient time has been excluded in Fig.~\ref{fig:cdhistories}. The definition of $C_D$ is given as
\begin{equation}
C_D=\frac{F_x}{\frac{1}{2}\rho U_{\infty}^2 A}
\label{eq:Cd}
\end{equation}
where $F_x$ is the streamwise component of the force acting on the disk directly calculated by integrating the pressure and viscous shear stress over the disk surface; $\rho$ is the density of the fluid; $A$ is the projected area of the disk in the streamwise direction. The time-averaged value and the r.m.s. value of $C_D$ are denoted as $\overline{C_D}$ and $C_{Drms}$, respectively. As shown in Fig.~\ref{fig:cdhistories}, significant discrepancies exist in the results of $C_D$ obtained from the five cases. The values of $\overline{C_D}$ for the full-disk, the half-disk, the quarter-disk, the eighth-disk and the 2D case are 1.054, 1.120, 1.038, 0.843 and 0.380, respectively. It appears that the $\overline{C_D}$ results of the first three cases agree well with the experimental data of $\overline{C_D}\approx1.1$ in \citep{roos1971some}. However, the $\overline{C_D}$ results of the eighth-disk and the 2D case are much lower, i.e., with the underestimations about $20\%$ and $64\%$, respectively, with respect to the result of the full-disk case. Furthermore, the r.m.s.~values of $C_D$ are also influenced by the simplification level of the simulations. As shown in Fig.~\ref{fig:cdhistories}, among the four 3D cases, the value of $C_{Drms}$ increases as the simplification level increases, i.e., the full-disk case has the minimum $C_{Drms}$ value. However, it should be noted that the uniform time history of $C_D$ for the 2D case indicates a steady flow pattern. It is easy to understand that the three-dimensionality of the flow is not resolved at all in the 2D case. For the flow around cylindrical structures, the 2D simulations are reported to be able to capture the unsteady vortex shedding pattern, see the example studies in Refs.~\citep{saha2007far,ong2009numerical,Tian2013208}. However, in the present 2D simulation for the axis-symmetric flow configuration, the unsteady vortex shedding pattern is greatly suppressed.


\begin{figure}
\begin{center}
\begin{tabular}{c}
\includegraphics[angle=0, width=0.7\columnwidth]{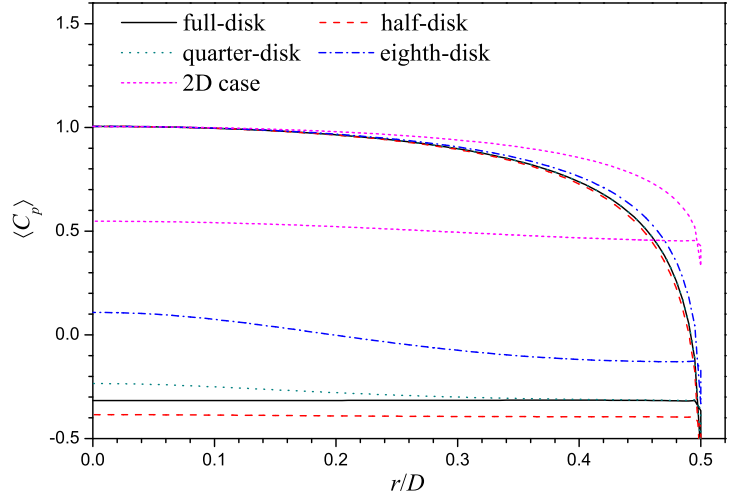}\\
\end{tabular}
\end{center}
\caption{Distributions of the time and azimuthal averaged pressure coefficient on the disk surface.}
\label{fig:CpMean}
\end{figure}

Fig.~\ref{fig:CpMean} shows the distributions of the time and azimuthal averaged pressure coefficient on the disk surface. The pressure coefficient $C_p$ is defined as
\begin{equation}
C_p=\frac{p-p_{\infty}}{\frac{1}{2}\rho U_{\infty}^2}
\label{eq:Cp}
\end{equation}
where $p$ is the local pressure, and $p_{\infty}$ is the reference pressure taken at the center of the inlet boundary. The time and azimuthal averaged value of $C_p$ is denoted as $\langle C_p\rangle$. For the flow normal to a circular disk, the drag force acting on the disk is dominated by the pressure distribution on the disk surface rather than the viscous shear component. Thus, the distribution of the mean pressure coefficient $\langle C_p\rangle$ is closely related to the mean drag coefficient $\overline{C_D}$. As shown in Fig.~\ref{fig:CpMean}, on the front side of the disk, the $\langle C_p\rangle$ distributions of all the 3D cases agree well with each other. On the back side of the disk, the $\langle C_p\rangle$ distributions of the first three 3D cases agree well with each other, while the fourth case (i.e., the eighth-disk case) has a lower mean pressure coefficient in absolute value compared with that of the first three 3D cases. Due to the seriously less resolved flow in the 2D case, the distribution of $\langle C_p\rangle$ on neither side of the disk is accurate, especially on the back side of the disk where a positive $\langle C_p\rangle$ value is observed. This is in conflict with the common sense that the pressure in the separated region of the bluff body should be negative, see examples in Refs.~\citep{narasimhamurthy2009numerical,xu2010large} as well as the present full-disk case. For the eighth-disk, a positive $\langle C_p\rangle$ value is observed in the region of $r/D<0.2$. It is shown that the flat $\langle C_p\rangle$ distribution on the back side of the disk is well captured in all the cases except for the eighth-disk case, just as those have been observed for the flow around other bluff bodies, see the examples for a circular cylinder in \citep{xu2010large} and for a plate with infinite span in \citep{najjar1998low}.


\begin{figure}
\begin{center}
\begin{tabular}{c}
\includegraphics[angle=0, width=0.7\columnwidth]{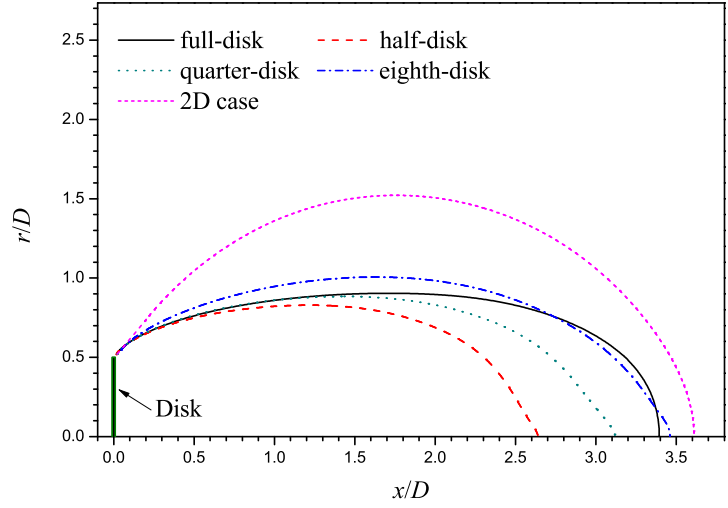}\\
\end{tabular}
\end{center}
\caption{Outlines of the mean recirculation bubble.}
\label{fig:BubbleOutline}
\end{figure}

Fig.~\ref{fig:BubbleOutline} shows the outlines of the time and azimuthal averaged recirculation bubble behind the disk in the five cases with different simplification levels. The outline of the mean recirculation bubble refers to the streamline connecting the separation point at the edge of the disk and the stagnation point (i.e., the location where the mean velocity equals to zero) in the wake. As shown in Fig.~\ref{fig:BubbleOutline}, the mean recirculation bubble of the 2D case is significantly wider and longer than those of the 3D cases. Although the results of $\overline{C_D}$ and the distributions of $\langle C_p\rangle$ for the first three cases are generally close to each other, the streamwise length of the mean recirculation bubble of these cases are very different. Besides the full-disk case, the size of the recirculation bubbles of the rest cases increases as the simplification level increases. Among the four simplified cases, the eighth-disk seems to give a better prediction of the shape of the mean recirculation bubble in the full-disk case.

\begin{figure}
\begin{center}
\begin{tabular}{c}
\includegraphics[angle=0, width=0.8\columnwidth]{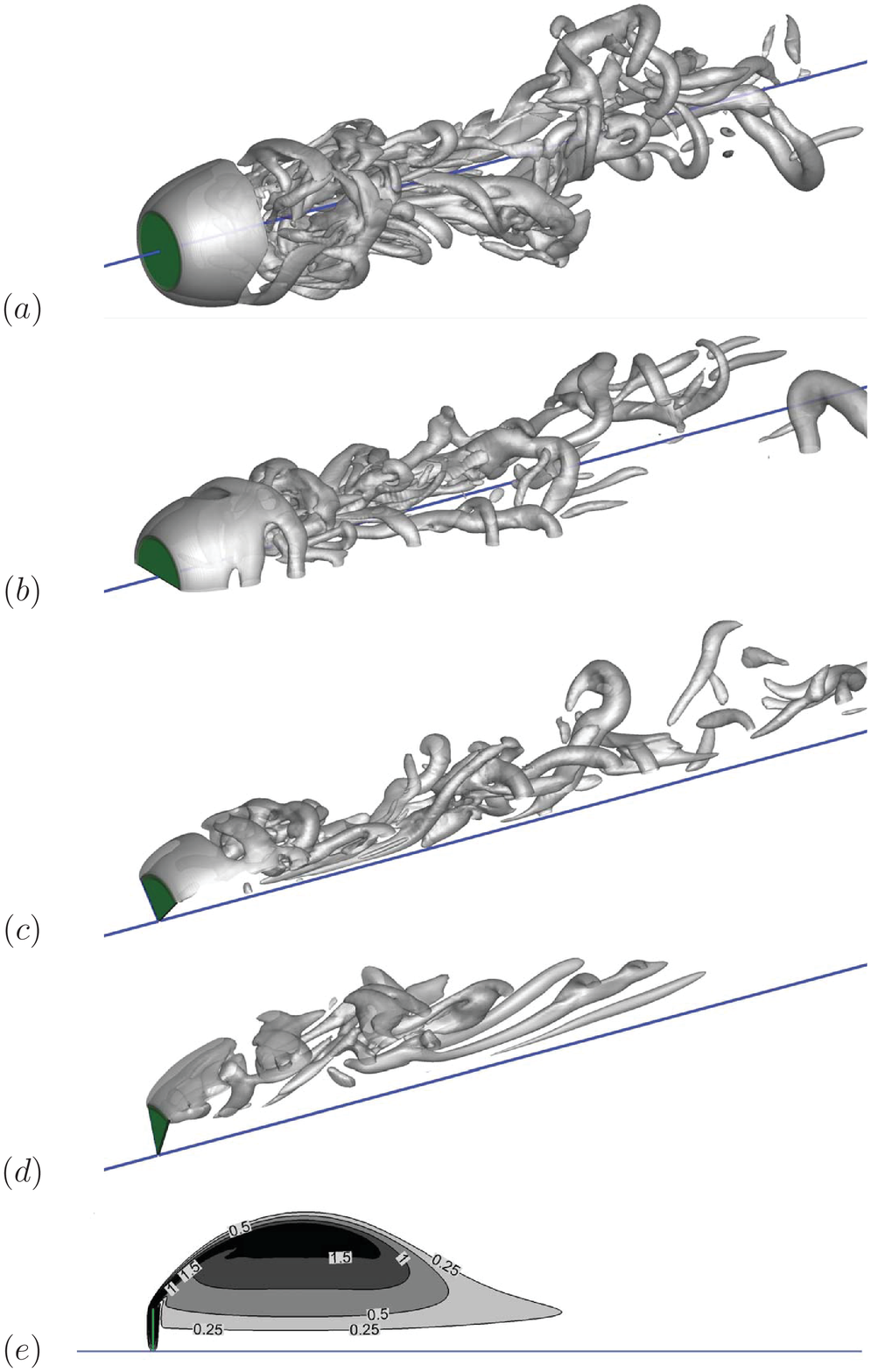}\\
\end{tabular}
\end{center}
\caption{Instantaneous snapshots of the wake flow for $(a)$ the full-disk, $(b)$ the half-disk, $(c)$ the quarter-disk, $(d)$ the eighth-disk and $(e)$ the 2D case.}
\label{fig:FlowView}
\end{figure}


Fig.~\ref{fig:FlowView} shows the instantaneous snapshots of the wake flow for the cases with different simplification levels. In Figs.~\ref{fig:FlowView}$(a$--$d)$, the 3D vortical structures are identified by the iso-surface of $Q=1$. The superiority of the $Q$-criterion to represent the 3D vortical topology is discussed in detail in \citep{hunt1988eddies}. In Fig.~\ref{fig:FlowView}$(e)$, the 2D steady flow is represented with the contours of the vorticity magnitude, denoted as $|\omega|$. The numbers along the contour lines in Fig.~\ref{fig:FlowView}$(e)$ refer to the value of non-dimensional vorticity magnitude $|\omega|D/U_{\infty}$. The green shade and solid blue line in Fig.~\ref{fig:FlowView} represent the disk and the centerline of the disk, respectively. As shown in Figs.~\ref{fig:FlowView}$(a$--$d)$, the 3D flow structure are captured in all the 3D simulations. Even there is only one eighth of the disk is simulated in the eighth-disk case, the vortical structures indicate a prominent three-dimensionality, see Fig.~\ref{fig:FlowView}$(d)$ . However, it is also observed that the 3D vortical structures for the half-disk, the quarter-disk and the eighth-disk have been confined more or less within the resolved domain. The evident discrepancies of the flow structures between the five cases result in the different characteristics of the $C_D$ time series and the mean flow quantities.


In summary, we proposed a numerical investigation on the effects of simplification levels of azimuthal extensions on the calculated results for the flow normal to a circular disk. Five cases, i.e., the full-disk, the half-disk, the quarter-disk, the eighth-disk and the 2D case, are carried out in total. Based on the comparisons of the results of $C_D$, ${C_{Drms}}$, $\langle C_p\rangle$, recirculation bubble and the wake flow pattern, it is concluded that the 3D effect plays an important roll in the characteristics of the wake flow behind a circular disk. Even the simulations based on the half-disk and the quarter-disk have provided good predictions of $\overline{C_D}$ and $\langle C_p\rangle$, the mean and the instantaneous flow characteristics have not been predicted well. Therefore, numerical simulations for the flow around a circular disk at a relatively high Reynolds number must be carried out with the 3D setup for the entire disk rather than for only one half of or a 2D plane of the disk.

\section*{Acknowledgment}

This work was supported by the Shanghai Yang Fan Program (Grant no.~15YF1406100) and the National Natural Science Foundation of China (Grant nos.~51509152 and 51239007). The simulations were performed on TianHe-1(A) at National Supercomputer Center in Tianjin, China.







\bibliographystyle{model2-names}
\bibliography{MyLiteratures}







\end{document}